\documentclass[letter,aps,prl,twocolumn,longbibliography,showpacs,reprint]{revtex4-1}
\usepackage[margin=0.8in]{geometry}
\usepackage{textcomp}
\usepackage{xfrac}
\usepackage{multirow}

\usepackage{natbib,graphicx,url,times,layouts,xcolor}
\usepackage{txfonts} 
\usepackage{cases} 
\definecolor{blueviolet}{rgb}{0.2, 0.2, 0.6}
\usepackage[pdftex, 
bookmarks=false, 
colorlinks=true,
allcolors=blueviolet,
pdfstartview={FitH}, 
]{hyperref}
\usepackage{pdfpages}
\usepackage{pgffor}

\newcommand{\reffig}[1]{Fig.~\ref{#1}}

\newcommand{\unit}{\ \mathrm}
\newcommand{\func}[2]{\mathrm{#1} \left( #2 \right)}

\renewcommand{\sin}[1]{\func{sin}{#1}}
\renewcommand{\exp}[1]{\func{exp}{#1}}
\renewcommand{\arg}[1]{\func{arg}{#1}}
\renewcommand{\cos}[1]{\func{cos}{#1}}

\newcommand{\cop}[1]{\hat{#1}^{\dag}}
\newcommand{\aop}[1]{\hat{#1}}
\newcommand{\copn}[2]{\hat{#1}^{\dag #2}}

\newcommand{\bra}[1]{\langle #1|}
\newcommand{\ket}[1]{|#1\rangle}

\begin{document}
\title{Cavity State Manipulation Using Photon-Number Selective Phase Gates}
\author{Reinier W. Heeres}
\author{Brian Vlastakis}
\author{Eric Holland}
\author{Stefan Krastanov}
\author{Victor V. Albert}
\author{Luigi Frunzio}
\author{Liang Jiang}
\author{Robert J. Schoelkopf}
\affiliation{Departments of Physics and Applied Physics, Yale University, New Haven, Connecticut 06520, USA}
\date{March 4, 2015}
\pacs{85.25.Hv, 03.67.Ac, 85.25.Cp}

\begin{abstract}
The large available Hilbert space and high coherence of cavity resonators makes these systems an interesting resource for storing encoded quantum bits. To perform a quantum gate on this encoded information, however, complex nonlinear operations must be applied to the many levels of the oscillator simultaneously. In this work, we introduce the Selective Number-dependent Arbitrary Phase (SNAP) gate, which imparts a different phase to each Fock state component using an off-resonantly coupled qubit. We show that the SNAP gate allows control over the quantum phases by correcting the unwanted phase evolution due to the Kerr effect. Furthermore, by combining the SNAP gate with oscillator displacements, we create a one-photon Fock state with high fidelity. Using just these two controls, one can construct arbitrary unitary operations, offering a scalable route to performing logical manipulations on oscillator-encoded qubits.
\end{abstract}

\maketitle

Traditional quantum information processing (QIP) schemes rely on coupling a large number of two-level systems (qubits) to solve quantum problems \cite{divincenzo1995quantum}. However, quantum information is fragile and susceptible to noise and decoherence processes, which implies the need for error-correction protocols. Many of these protocols have been proposed \cite{nielsen2010quantum}, yet all require significant resources by constructing a single logical qubit consisting of many physical qubits. A drawback of this approach is that logical gate operations become more complicated because multiple physical systems need to be addressed. Alternatively, one could encode one bit of quantum information within a higher-dimensional Hilbert space, such as a propagating mode (i.e. coherent light) or a stationary mode in a cavity \cite{braunstein2005quantum}. Several proposals exist to use this larger Hilbert space for redundant encoding to allow quantum error correction. Some of these are very similar to multi-qubit based codes \cite{gottesman2001encoding,cafaro2012quantum} and others are based on superpositions of coherent states \cite{leghtas2013hardware,mirrahimi2014dynamically}, or so-called cat-codes.

A major complication for quantum computation with harmonic oscillators, however, is their linear energy spectrum and thus degenerate transitions, making quantum control difficult. This problem can be resolved by coupling the harmonic oscillator to a qubit, giving the system a controllable nonlinearity. In the case of a qubit that can be tuned into resonance with an oscillator, one can create arbitrary quantum states by swapping excitations from the qubit to the cavity mode \cite{law1996arbitrary,hofheinz2008generation,hofheinz2009synthesizing}. However, there are several advantages of having a far off-resonantly, or dispersively, coupled qubit instead of a resonant one. Full microwave control without the requirement of a flux-bias line is the most important factor experimentally, as it decreases the sensitivity to flux noise and improves scalability.

It has been shown that universal control is possible using a single non-linear term in addition to linear controls \cite{lloyd1999quantum}, but no efficient construction has been provided and so far there have been no experimental demonstrations. Using the dispersive nonlinearity, some special operations can be realized efficiently, for example mapping an arbitrary qubit state to a superposition of coherent states in a cavity \cite{vlastakis2013deterministically}. Here we extend the set of operations in the dispersive regime with the Selective Number-dependent Arbitrary Phase or SNAP gate, which manipulates the phase of one or more photon-number states. We also show that photon-number populations can be controlled by combining the SNAP gate with ordinary displacements of the oscillator. In a related work we prove that these tools provide arbitrary control and describe a constructive method to generate unitary operations \cite{stefan2015universal}.

\begin{figure*}[thbp]
\begin{center}
    \mbox{
      \includegraphics{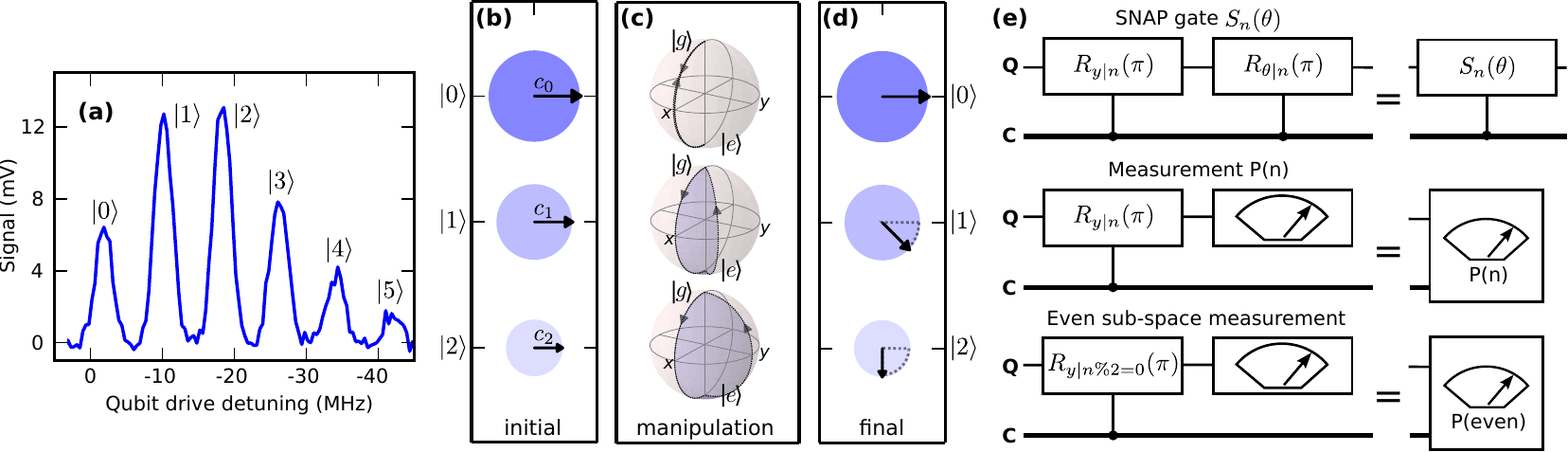}
    }
\end{center}
\caption{\label{fig:fig1} \textbf{Qubit spectroscopy and geometric phase gate. (a)} Qubit spectrum with storage cavity population $\bar{n} \approx 2$, showing that the qubit transition frequency depends on the number of photons in the cavity. \textbf{(b)} Phasor representation of a cavity state: the arrow $c_n$ corresponds to the complex amplitude of state $\ket{n}$; the area of the circle is proportional to $|c_n|^2 = p(n).$ \textbf{(c)} An example of the SNAP gate: two $\pi$ pulses on the qubit along different axes result in a trajectory that encloses a geometric phase and allows a rotation on each $c_n$ selectively. \textbf{(d)} Final state after operation in (c), which results in a controllable phase evolution on each $c_n$. \textbf{(e)} Quantum circuit representation of the gates used in this work, where \textbf{Q} and \textbf{C} correspond to the qubit and the cavity state respectively. $R_{y|n}(\phi)$ should be read as ``a rotation by angle $\phi$ around $y$ conditional on $n$ photons in the cavity''; ``\%'' is the modulo operator.
}
\end{figure*}

We use a circuit quantum electrodynamics architecture (cQED) \cite{haroche2013exploring,deleglise2008reconstruction} with two cavities coupled to one superconducting transmon qubit \cite{blais2004cavity,wallraff2004strong}. One of the cavities is used as a long-lived storage cavity, the other as a fast readout resonator. The latter is used as an ancillary system and will not be included in our description of the quantum state; from here on, the word cavity will refer to the storage cavity. Details of the sample and measurement setup are described in \cite{vlastakis2013deterministically} and \cite{heeres2015sup}. When the qubit-cavity detuning is much larger than their coupling strength, the system can be described using the dispersive Hamiltonian \cite{schuster2007resolving}:
\begin{equation}
 \sfrac{H}{\hbar} = \omega_c \cop{a}\aop{a} + \omega_q \ket{e}\bra{e} + \chi \cop{a}\aop{a}\ket{e}\bra{e}
\end{equation}
with $\omega_c$ and $\omega_q$ the cavity and qubit transition frequency respectively, $\aop{a}$ the annihilation operator of a cavity excitation, $\chi$ the dispersive shift and $\ket{e}\bra{e}$ the qubit excited-state projector. Higher-order terms are usually ignored, but in this work we quantify the Kerr term $\frac{K}{2}\copn{a}{2}\aop{a}^2$ (see \cite{kirchmair2013observation}), the corrections to the dispersive shift $\frac{\chi'}{2}\copn{a}{2}\aop{a}^2\ket{e}\bra{e}$, $\frac{\chi''}{6}\copn{a}{3}\aop{a}^3\ket{e}\bra{e}$ and the correction to Kerr $\frac{K'}{6}\copn{a}{3}\aop{a}^3$.

We are operating in the number-split regime \cite{schuster2007resolving,gambetta2006qubit,johnson2010quantum} where the qubit frequency shift per photon in the cavity, $\chi$, is larger than both the qubit and cavity transition linewidths. This results in a spectrum as shown in \reffig{fig:fig1}a and means that the qubit can be addressed selectively if and only if there are $n$ photons in the cavity. It is natural to represent the cavity state in the Fock basis $\ket{\psi_c} = \sum_{n=0}^{\infty} c_n \ket{n}$. We can depict this state as a set of phasors with length $|c_n| = \sqrt{p(n)}$, where $p(n)$ is the probability that the cavity is populated by $n$ photons, and $\arg{c_n} = \theta_n$ the associated quantum phase. \reffig{fig:fig1}b shows an example of this graphical way to represent a pure state. The area of the colored circle can be directly interpreted as the probability $p(n)$. The full system state is given by $\ket{\psi} = \sum \ket{\psi_c} \otimes \ket{\psi_q}$, and we use the qubit as an ancilla to manipulate the state of the cavity.

The SNAP gate we propose here consists of a geometric phase, similar to the Berry phase \cite{leek2007observation}, applied selectively to the $n$-th Fock-state component by doing two rotations on the qubit, using a weak drive (Rabi frequency $\Omega \ll \chi$) at frequency $\omega_d = \omega_q + n\chi$. If the two pulses are performed along different axes, the trajectory on the Bloch sphere encloses a solid angle, which corresponds to the acquired geometric phase. The qubit itself should start and end in the ground state, effectively disentangling the qubit and cavity after each operation, such that undesired effects due to qubit relaxation and decoherence are minimized before and after the SNAP gate.
When the drive strengths are small, we can superpose drives on many different Fock state components into a final control pulse that applies an arbitrary quantum phase to each coefficient $c_n$, as indicated in \reffig{fig:fig1}c. We will also use the selective drives to measure whether there are $n$ photons in the cavity, or, by applying a selective pulse on all the even or the odd peaks, to measure the parity of the cavity state. \reffig{fig:fig1}e shows these operations in quantum circuit notation. Mathematically the SNAP gate $S_n(\theta)$ on a single Fock state $\ket{n}$ can be described as
\begin{equation}
  S_n(\theta) = e^{i\theta \ket{n}\bra{n}}
\end{equation}
and the generalized SNAP gate $S(\vec{\theta})$ acting on multiple Fock state components as:
\begin{equation}
  S(\vec{\theta}) = \prod_{n=0}^{\infty} S_n(\theta_n)
\end{equation}
with $\vec{\theta} = \{\theta_n\}_{n=0}^{\infty}$.

Several well-known operations are straight-forward to express in this form, for example a rotation by $\phi$ in the IQ-plane: $\theta_n = n\phi$, i.e. a phase linearly increasing with photon number $n$. Another example described by a simple phase relation is Kerr-evolution, given by $\theta_n \propto n^2$. However, more exotic operations can also be performed using this gate. Consider the phase pattern $\vec{\theta} = \{0,\phi,0,\phi,\dots\}$, giving an equal phase to only the odd photon-number components. When performed on a cavity state $\ket{\beta}$, this corresponds to a coherent rotation to $\cos{\sfrac{\phi}{2}}\ket{\beta} - i\sin{\sfrac{\phi}{2}}\ket{{-\beta}}$, as shown by simulated Wigner functions in the supplemental material \cite{heeres2015sup}. When using a cat-code \cite{mirrahimi2014dynamically} with $\ket{\beta}$ and $\ket{{-\beta}}$ as the logical basis states, this manipulation is a rotation on the encoded qubit.

\begin{figure*}[thbp]
\begin{center}
  \mbox{\includegraphics{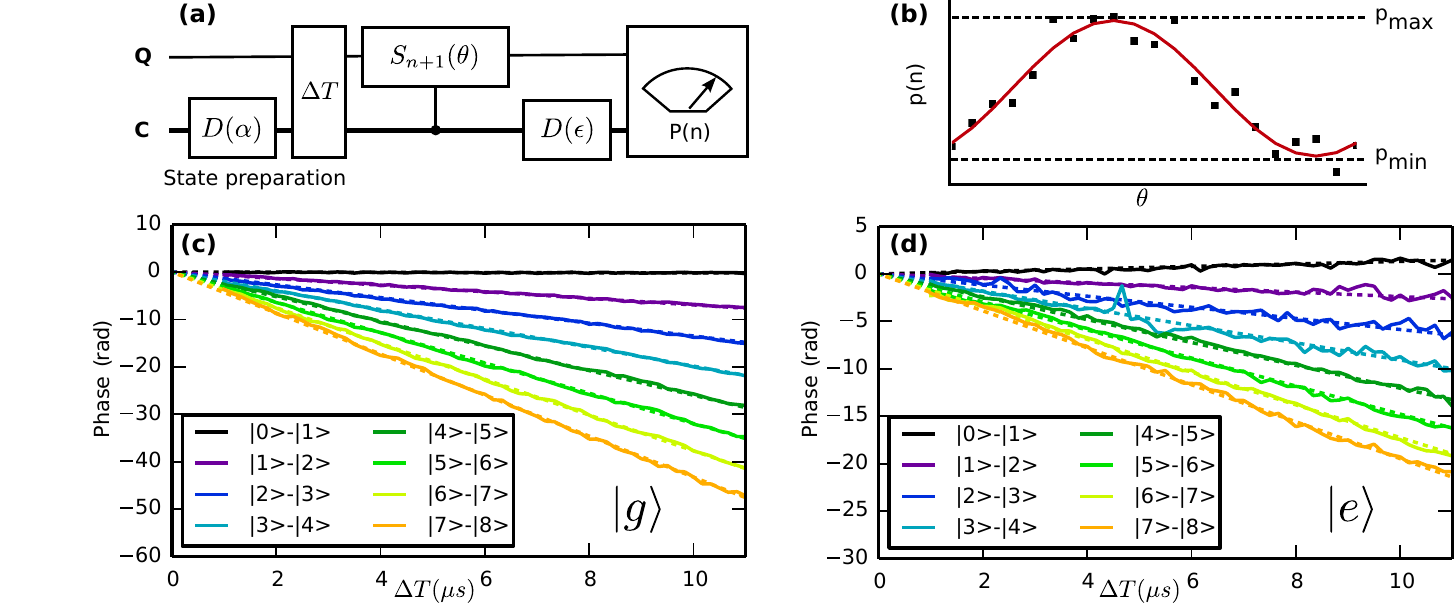}}
\end{center}
\caption{\label{fig:phase_meas} \textbf{Phase evolution measurement. (a)} Quantum circuit describing the pulse sequence: state preparation consists of an optional $\pi$ pulse on the qubit (not shown), a displacement of the cavity $D(\alpha)$ and a variable waiting time $\Delta T$. The phase between $c_{n}$ and $c_{n+1}$ is determined by varying the applied quantum phase using the SNAP gate $S_{n+1}(\theta)$, displacing by $\epsilon$ and measuring the population $p(n)$. \textbf{(b)} Typical measurement signal $p(n)$ versus applied phase $\theta$ at fixed $\Delta T$: a sine wave with minimum $p_{\mathrm{min}}$ and maximum $p_{\mathrm{max}}$. The phase between $c_{n}$ and $c_{n+1}$ corresponds to the case of destructive interference and is determined from the fit. \textbf{(c)} Phase evolution with qubit in the ground state. Each trace is the (unwrapped) measured phase difference between neighboring Fock-state components versus waiting time $\Delta T$. The rotating frame is chosen such that there the phase difference between $\ket{0}$ and $\ket{1}$ is close to zero. From the fit we extract drive detuning $\sfrac{\Delta}{2\pi} = (-1.1 \pm 0.7) \unit{kHz}$ and Hamiltonian parameters $\sfrac{K}{2\pi} = (-107.9 \pm 0.5) \unit{kHz}$ and $\sfrac{K'}{2\pi} = (3.4 \pm 0.1) \unit{kHz}$. \textbf{(d)} Same as in (c) but with the qubit starting in the excited state. The frame on the cavity generator is adjusted by $-8300.0 \unit{kHz}$ and from the fit we find $\sfrac{\Delta}{2\pi} = (17.6 \pm 0.9) \unit{kHz}$ to give $\sfrac{\chi}{2\pi} = (-8281.3 \pm 1) \unit{kHz}$, $\sfrac{\chi'}{2\pi} = (48.8 \pm 0.8) \unit{kHz}$ and $\sfrac{\chi''}{2\pi} = (0.5 \pm 0.2) \unit{kHz}$.
}
\end{figure*}

The SNAP gate provides a natural way to determine the relative phase between Fock states. To measure these phases we perform an interference experiment between neighboring number states facilitated by a small displacement. This operation effectively allows us to map a relative phase between number state $\ket{n}$ and $\ket{n+1}$ to a change in probability of finding the system in $\ket{n}$. In the limit of a small real displacement amplitude $\epsilon \ll \sfrac{1}{\sqrt{n_{max}}}$, with $n_{max}$ the maximum photon number considered, the displacement operator can be approximated by its first order expansion:
\begin{equation}
D(\epsilon) = \exp{\epsilon\cop{a} - \epsilon^*\aop{a}} \approx 1 + \epsilon\cop{a} - \epsilon^*\aop{a}
\end{equation}
In this limit only neighboring Fock-state components interact. We now consider the effect on $c_{n}$ after a displacement $D(\epsilon)$ with real $\epsilon$. Depending on the phase difference $\phi = \arg{c^*_{n+1}c_{n}}$ the coefficients $c_{n+1}$ and $c_n$ will add constructively or destructively, corresponding respectively to an increased or decreased probability of finding $n$ photons. By applying the SNAP gate $S_{n+1}(\theta)$ to add phase $\theta$ to $c_{n+1}$ before the small displacement, we can find the phase $\theta_{n+1,n}$ which aligns the vectors $c_{n+1}$ and $c_n$. We have used this scheme to measure the phase difference between neighboring Fock-states as a function of time, i.e. the system evolution under influence of the system's Hamiltonian, both with the qubit starting in the ground and in the excited state. The results are shown in \reffig{fig:phase_meas} and allow us to extract all relevant Hamiltonian parameters directly.

\begin{figure}[thbp]
\begin{center}
\mbox{\includegraphics{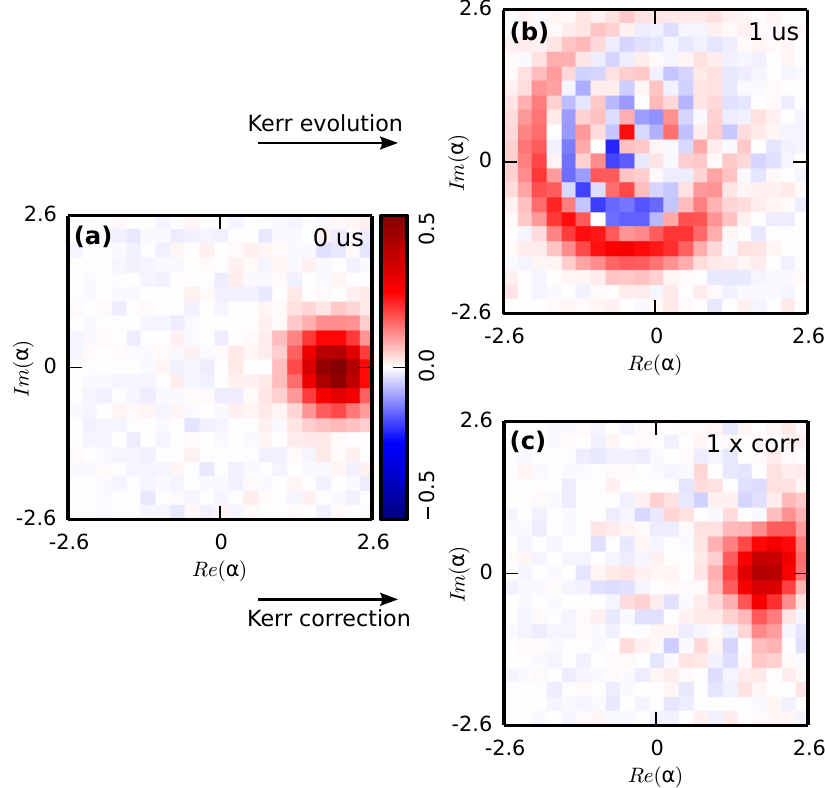}}
\end{center}
\caption{\label{fig:kerr_corr} \textbf{Kerr correction. (a)} Wigner function of the cavity in a coherent state directly after a displacement operation $D(2)$. \textbf{(b)} Wigner function after $1 \unit{\mu s}$ of free evolution dominated by the Kerr effect. \textbf{(c)} Wigner function after 1 step of the Kerr-correction scheme (taking $1 \unit{\mu s}$). The fidelity to a coherent state is $F = 0.97$ (see \cite{heeres2015sup}). 
}
\end{figure}

Another application of the SNAP gate is to compensate the deterministic phase evolution due to the Kerr effect \cite{kirchmair2013observation}, which is typically an undesired interaction. We have used a SNAP gate to cancel this phase periodically by applying an operation $S(\vec{\theta})$ that exactly cancels the phases acquired in the $1 \unit{\mu s}$ duration of the operation. The pulse consists of superposed sideband-modulated Gaussians with $\sigma = 125 \unit{ns}$, corresponding to a spectral width $\sigma_f = \sfrac{1}{2\pi\sigma} \approx 1.25 \unit{MHz}$ and drives the first 11 photon-number resolved peaks simultaneously. The result of a single step of Kerr-cancellation on a coherent state $\ket{\beta = 2}$ is shown in \reffig{fig:kerr_corr}, with \reffig{fig:kerr_corr}b showing the cavity Wigner function when no Kerr-correction pulse is applied and \reffig{fig:kerr_corr}c showing the corrected cavity state. The supplemental material \cite{heeres2015sup} contains Wigner functions showing the free-evolution as well as the Kerr-corrected state from $1$ to $14 \unit{\mu s}$, i.e. with up to 14 Kerr-cancellation pulses applied sequentially. The fidelity of a single correction step is limited due to qubit relaxation and dephasing to about $0.96$.

\begin{figure}[tp]
\begin{center}
  \mbox{\hspace{-1cm}\includegraphics{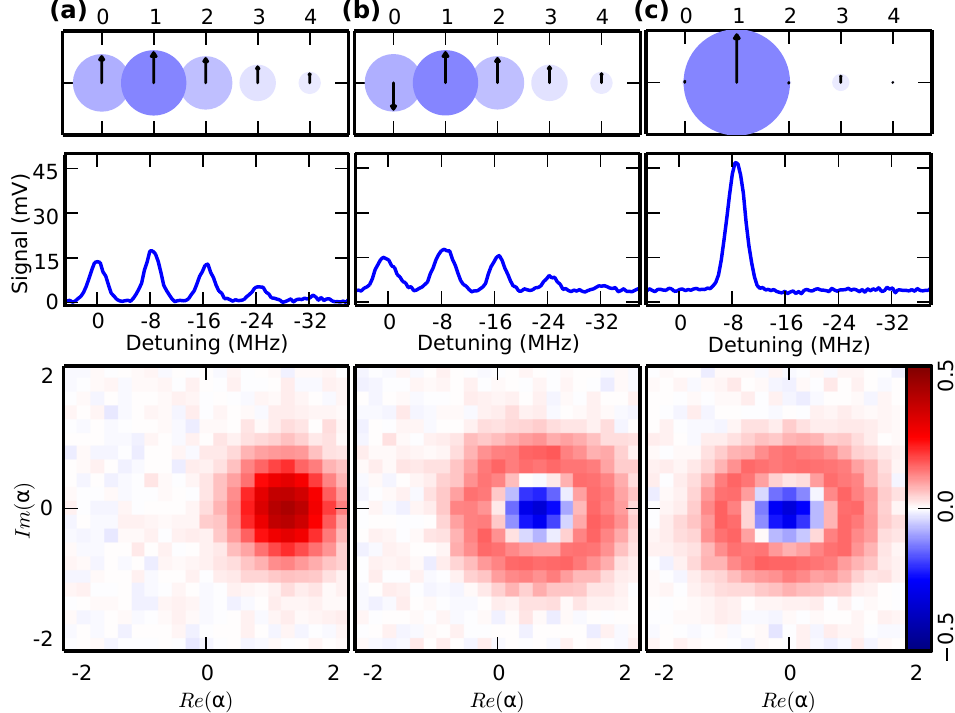}}
\end{center}
\caption{\label{fig:fock_create} \textbf{Fock state creation.} Phasor representation, qubit spectrum and Wigner function after each of the steps in the Fock state creation scheme. \textbf{(a)} Displacement $\beta_1 = 1.14$, \textbf{(b)} Applying a $\pi$ phase shift on $\ket{0}$ as well as cancelling Kerr induced phases on other components. \textbf{(c)} Final displacement $\beta_2 = -0.58$ to complete Fock state $\ket{1}$ creation.
}
\end{figure}

Combining the SNAP gate with cavity displacements allows us to not only manipulate the quantum phases, but also the Fock state populations. As an example we show in \reffig{fig:fock_create} that we can create the Fock state $\ket{1}$ by applying the operation $D(\beta_2)S(\vec{\theta})D(\beta_1)$. The displacements in this sequence are numerically optimized while $\vec{\theta}$ is fixed to be $(\pi,0,0,...)$. The displacement by $\beta_1$ populates the cavity with a coherent state (\reffig{fig:fock_create}a). After performing a phase flip on the $\ket{0}$ component using a SNAP gate which also corrects for Kerr-evolution (\reffig{fig:fock_create}b), the Wigner function corresponds to a displaced Fock state \cite{de1990properties}. When performing qubit spectroscopy, however, a Poisson photon-number distribution is still present. The second displacement (\reffig{fig:fock_create}c) simply translates the Wigner function back to the origin, but in the Fock-basis the effect is pronouncedly different: qubit spectroscopy reveals that the state is converted to the Fock state $\ket{1}$. Simulations show that the sequence should give a fidelity $F = 0.98$, whereas we find $F = 0.90$ after density matrix reconstruction \cite{smolin2012efficient} based on the experimental data (see \cite{heeres2015sup}), which we attribute to qubit decoherence.

A simple extension of the displacement - phase - displacement protocol used to create the $\ket{1}$ state allows the construction of operations to climb the ladder of Fock states by successive application, an indication that arbitrary state preparation is possible. In fact, the SNAP gate and displacements together provide universal control, allowing any unitary operator to be constructed \cite{stefan2015universal}. With $n_{\mathrm{max}}$ the maximum photon number, state preparation can be achieved in $O(n_{\textrm{max}})$ operations, and arbitrary unitaries in $O(n_{\mathrm{max}}^2)$ operations. This sets an upper bound for the speed at which the large Hilbert space available in oscillators can be manipulated.

In summary, we have introduced the SNAP gate, which allows to manipulate a cavity state by applying a controlled, arbitrary phase to each individual Fock state component. We have used the SNAP gate to determine all relevant Hamiltonian parameters by measuring the evolution of the relative phase difference between neighboring Fock states in an interference experiment. Two useful cavity operations have also been presented: correcting for evolution due to the Kerr effect and creating a one photon Fock state deterministically. Using the extensions in \cite{stefan2015universal}, our scheme could potentially be used to manipulate quantum information encoded in an oscillator.

We would like to thank Mazyar Mirrahimi, Michel H. Devoret and Wolfgang Pfaff for helpful discussions. This work was partially supported by the Army Research Office (ARO) grant W911NF-14-1-0011, AFOSR MURI, the Alfred P. Sloan Foundation, the Packard Foundation, the Yale QIMP Fellowship (R.W.H.), NSF PHY-1309996 (B.V.) and the NSF Graduate Research Fellowship Program under Grant DGE-1122492 (V.V.A.).

\bibliographystyle{apsrev4-1}
\bibliography{transmons_abbrv}

\newpage\newpage


\section*{Supplementary material}

# Supplemental material for "Cavity State Manipulation Using Photon-Number Selective Phase Gates"

Reinier W. Heeres, Brian Vlastakis, Eric Holland, Stefan Krastanov,
Victor V. Albert, Luigi Frunzio, Liang Jiang, and Robert J. Schoelkopf
*Departments of Physics and Applied Physics, Yale University, New Haven, Connecticut 06520, USA*
(Dated: March 4, 2015)

### System and parameters

The sample is a a vertical transmon qubit in a high-purity aluminium cavity, measured in a dilution refrigerator at $\sim 20$ mK base temperature. Both the sample and the setup are identical to the ones described in [2] of the main text. We use high-power readout [1] to measure the state of the qubit.

The qubit relaxation time is $T_{1q} = 21$ μs, and the Ramsey time $T_{2q} = 30$ μs. Because the qubit resides in the excited state for about 50% of the SNAP gate time $T_g = 1$ μs, the expected gate fidelity is therefore approximately $F_g \approx e^{-T_g/2T_{1q}} e^{-T_g/2T_{2q}} = 0.96$. The storage cavity lifetime is $T_{1c} \sim 50$ μs.

Further parameters are specified in table I. Note that Kerr ($K$) and the next order correction ($K'$) agree within the error bars when measured with the two different methods. For the dispersive shift ($\chi$), however, we see a small discrepancy which we don't fully understand.

| Term | Phase evolution | Value 2 | Units |
|---|---|---|---|
| $\omega_q/2\pi$ | – | 7627.05 | MHz |
| $\omega_c/2\pi$ | $8226.787 \pm 0.001$ | – | MHz |
| $\omega_{ro}/2\pi$ | – | 9376.75 | MHz |
| $K/2\pi$ | $-107.9 \pm 0.5$ | $-106.2$ | kHz |
| $K'/2\pi$ | $3.4 \pm 0.1$ | $3.4$ | kHz |
| $\chi/2\pi$ | $-8281.3 \pm 1$ | $-8273 \pm 25$ | kHz |
| $\chi'/2\pi$ | $48.8 \pm 0.8$ | $60.9 \pm 1.6$ | kHz |
| $\chi''/2\pi$ | $0.5 \pm 0.2$ | – | kHz |

TABLE I. **System parameters.** Measured system parameters; $\omega_{ro}$ is the frequency of the readout cavity. The "Phase evolution" column contains data from the measurements in Fig. 2 of the main text. The colum "Value 2" lists values from spectroscopy (Fig. S1) and free evolution Wigner functions Fig. S3.

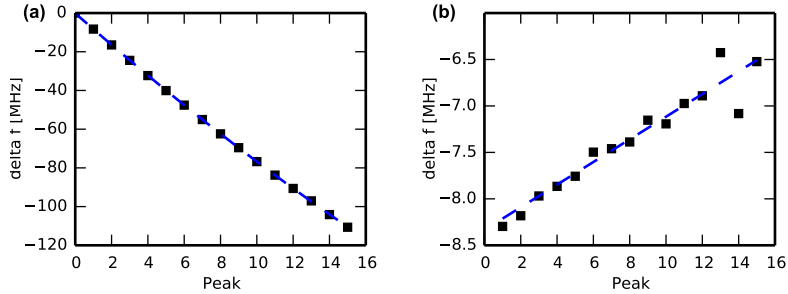

FIG. S1. **Photon-number dependence of qubit frequency. a.** Qubit transition frequency versus number of photons in the storage cavity, measured by performing spectroscopy using a slow $\pi$ pulse after a displacement of the cavity (as in Fig. 1a of the main text). The fit finds $\chi = (-8273 \pm 25)$ kHz and $\chi' = (60.9 \pm 1.6)$ kHz. **b.** Same data as in (a), but now the difference between neighboring peaks is plotted. The fact that the difference is modeled well by a straight line suggests that higher order corrections ($\chi''$) are small and can be neglected.

## Phase measurement analysis

In the free evolution phase measurement using the SNAP gate in the main text, we measure the phase difference $\phi_{n+1,n} = \arg\left(c_{n+1}^* c_n\right)$ between states $|n+1\rangle$ and $|n\rangle$. This evolves linearly with the energy difference, so $\phi_{n+1,n}(t) = (E(n+1) - E(n))t$. We should therefore look at the difference $H(n+1) - H(n)$, with the Hamiltonian Eq. 1 of the main text, including the higher order corrections. Evaluating this, we find that the phase difference evolves as:

$$\phi_{n+1,n}(t) = \Delta t + (K - \frac{K'}{2})nt + \frac{K'}{2}n^2 t \tag{1}$$

When the qubit is in the excited state the system evolves identically, but with $\Delta_{|e\rangle} = \Delta + \chi$, $K_{|e\rangle} = K + \chi'$ and $K'_{|e\rangle} = K' + \chi''$.

## Additional simulations and data

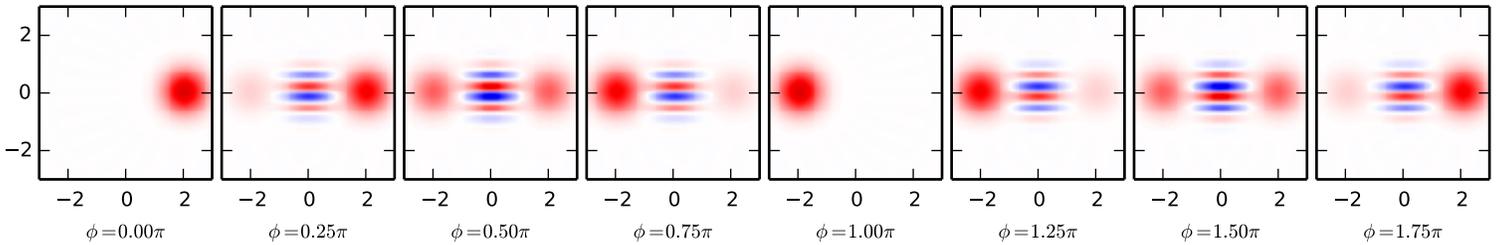

FIG. S2. **Simulated example of a non-trivial SNAP gate operation.** Simulated Wigner functions taken after the operation $S(\vec{\theta})$, with $\vec{\theta} = (0, e^{i\phi}, 0, e^{i\phi}, \ldots)$, for several different values of $\phi$. In an encoding where $|\beta\rangle$ and $|-\beta\rangle$ are the encoded logical states, this operation corresponds to a coherent rotation. It takes the initial state $|\beta\rangle$ through $|\beta\rangle - i|-\beta\rangle$ to $|-\beta\rangle$. Note that the operation does not change photon number and therefore preserves parity, as visible by the origin of the Wigner function remaining 0.

---

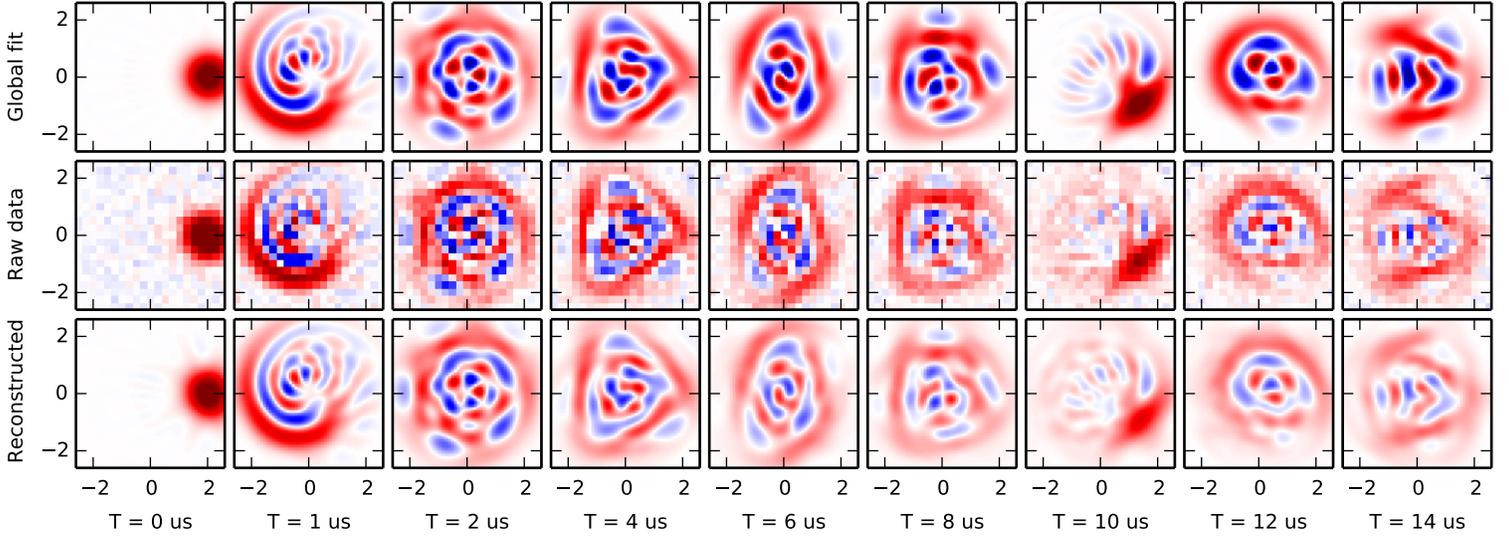

FIG. S3. **Reconstruction of Kerr data.** A second method to determine Hamiltonian parameters is by taking Wigner functions after different times T of free evolution. The middle row shows the raw data, the lower row shows Wigner functions of the reconstructed density matrices. The top row is a global fit of a model with three free parameters to the reconstructed wigner functions. From the fit we find $\Delta = -4.1$ kHz, kerr $K = -106.2$ kHz and kerr prime $K' = 3.4$ kHz.

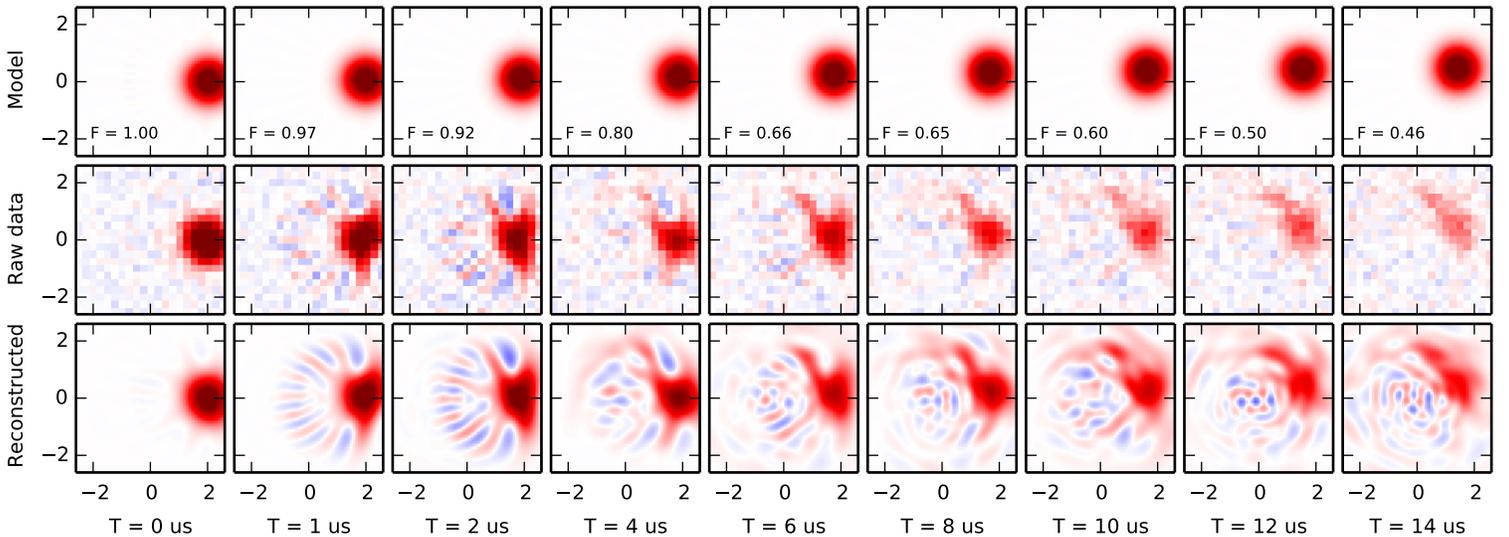

FIG. S4. **Reconstruction of Kerr-correction data.** To quantify how well we can perform a single step of Kerr cancellation, we take Wigner functions after $n$ steps of the Kerr-correction protocol. Each step is a single SNAP gate which takes 1 $\mu$s. The middle row is the raw data, the lower row shows Wigner functions of the reconstructed density matrices. The top row is a global fit to an ideal model of a coherent state with a detuning and corrected for cavity decay ($|\beta e^{i2\pi\Delta t} e^{-t/T_1}\rangle$). For the two free fit-parameters we find $\Delta = 3.6$ kHz and cavity $T_1 = 47.9$ $\mu$s. The numbers $F$ correspond to the overlap of the model with the reconstructed density matrices ($\text{Tr}\,(\rho_m \rho_r)$).

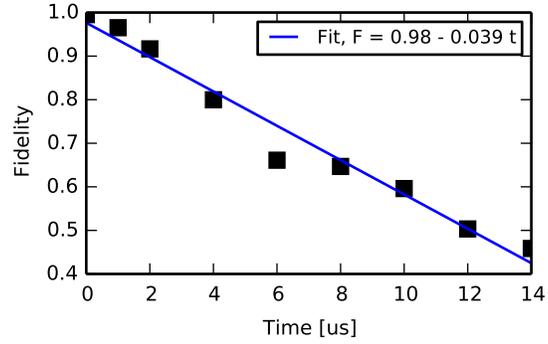

FIG. S5. **Fidelity versus number of corrections.** This figure shows the fidelity of the Kerr-corrected state to a coherent state model versus number of corrections (the number $F$ from the the fits in Fig. S4). Although in principle the fidelity should be an exponentially decaying function of $n$, we can still approximate it using a linear fit in this regime and find that the infidelity $1 - F$ of each step is about 0.04.

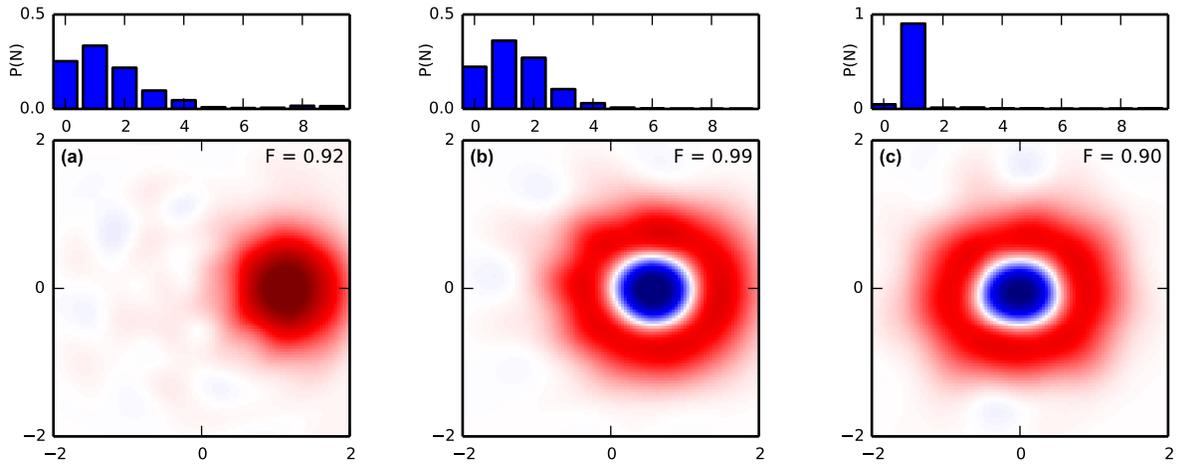

FIG. S6. **Reconstruction of Fock-state creation data.** Wigner functions of the density matrices reconstructed from the data of Fig. 4 in the main text. The indicated fidelity F is defined as $Tr\{\rho_r \rho_t\}$ (a.) After the initial displacement $\beta_1 = 1.14$. **b.** After applying the SNAP gate; the target for fidelity is $D(0.58)|1\rangle$. **c.** After the final displacement $\beta_2 = -0.58$.



\end{document}